# Study of the Pioneer Anomaly: A scientific detective story

Viktor T. Toth

*Ottawa, Ontario, K1N 9H5, Canada*

vttoth@vttoth.com

*Abstract* – **NASA's first two deep space missions, Pioneers 10 and 11, have been travelling through the outer solar system for three decades. A slight deviation from their calculated trajectories presents an as yet unsolved scientific mystery. The use of recently recovered Doppler and telemetry data may help us develop a better understanding of this anomaly, and decide whether or not it is due to a force of on-board origin.**

## I. INTRODUCTION

On April 27, 2002 DSS-63, the 70 meter antenna at the Madrid tracking station of NASA's Deep Space Network picked up an artificial radio signal from deep space. It was a signal sent almost 12 hours earlier by Pioneer 10, the first probe ever to leave the inner solar system and travel beyond the orbit of Mars. The signal was incredibly weak, only about –180 dBm (or about $10^{-21}$ Watts) but it was strong enough to extract useful data: although barely, but the spacecraft was still functional.

It was more than 30 years earlier, on March 2, 1972 that the night sky lit up over Cape Canaveral and Pioneer 10 began its perilous journey. Originally designed for an 18-24 month mission to travel beyond the orbit of Mars, cross the asteroid belt, and fly by Jupiter, Pioneer 10 surpassed the wildest expectations of its designers, and remained functional for more than 30 years.

The transmission on April 27, 2002 was the last[*] we heard from humanity's first probe to enter interstellar space. Pioneer 10 is now silent, but it left behind a mystery that keeps researchers busy. Simply put, Pioneer 10 is not where it should be if its motion was governed solely by the known laws of celestial mechanics.

The implications are tremendous: deep space probes like Pioneer 10 may provide a means to use the solar system as a laboratory to test gravity physics beyond Einstein's predictions. However, the possibility exists that the anomaly is a mere artefact, a result of our incomplete understanding of the known laws of physics and, in particular, of the engineering of the probe itself. To improve that understanding, we have been forced to dig into old records and decades old data sets as we strive to build a more complete profile of the Pioneer missions.

## II. THE PIONEER MISSIONS

Pioneers 10 and 11 (Figure 1) were the first probes designed to explore the outer solar system. Both spacecraft flew by Jupiter and made close-up observations of the gas giant. Pioneer 10 then followed a hyperbolic escape trajectory out of the solar system. Pioneer 11 was retargeted for a second flyby, several years later, of the ringed planet Saturn, after which it also began a never-ending journey escaping the solar system, in a direction roughly opposite that of Pioneer 10.

Pioneers 10 and 11 were very simple in design, and were dominated in appearance by a large parabolic antenna that was designed to maintain communication with Earth-based tracking stations even across distances measured in light-hours. To ensure that the spacecraft had adequate electrical power far from the Sun, nuclear power sources (radioisotope thermoelectric generators, or RTGs) were used on board [2]. These generators convert the heat produced by the spontaneous fission of nuclear fuel ($^{238}$Pu) into electricity using bimetallic thermocouples. The conversion is inefficient; much (~94%) of the heat produced by the RTGs is waste heat that is radiated into space.

The Pioneer spacecraft are "spin-stabilized": they are spinning at several revolutions per minute. Conservation of angular momentum dictates that, absent external forces, the spacecraft's spin axis continues to point in the same direction. The spin axis coincides with the axis of the antenna, and the spacecraft is oriented such that the antenna points at the Earth. This way, communication with the Earth was maintained. From time to time, it was necessary to adjust the spin axis to track the Earth's apparent motion across the sky as seen from the spacecraft; this was accomplished by so-called precession manoeuvres. Between such manoeuvres the spacecraft is flying undisturbed.

It was recognized early on that a spacecraft that is flying undisturbed by manoeuvres for months at a time may serve as an excellent, highly sensitive platform to conduct gravitational experiments. Pioneers 10 and 11 were at one time thought to be useful as a means to search for an hypothetical tenth planet, and also for gravity waves of extrasolar origin. While these searches were ultimately not fruitful, they yielded an unexpected surprise: a small but persistent deviation between the calculated and actual trajectories of both spacecraft [3].

---

[*] A year later, in January 2003, another attempt was made to contact Pioneer 10; although a weak signal was detected, it was not strong enough for the receiver to lock and no data was extracted. Subsequent attempts to communicate with the spacecraft, the last in early 2006, were not successful.

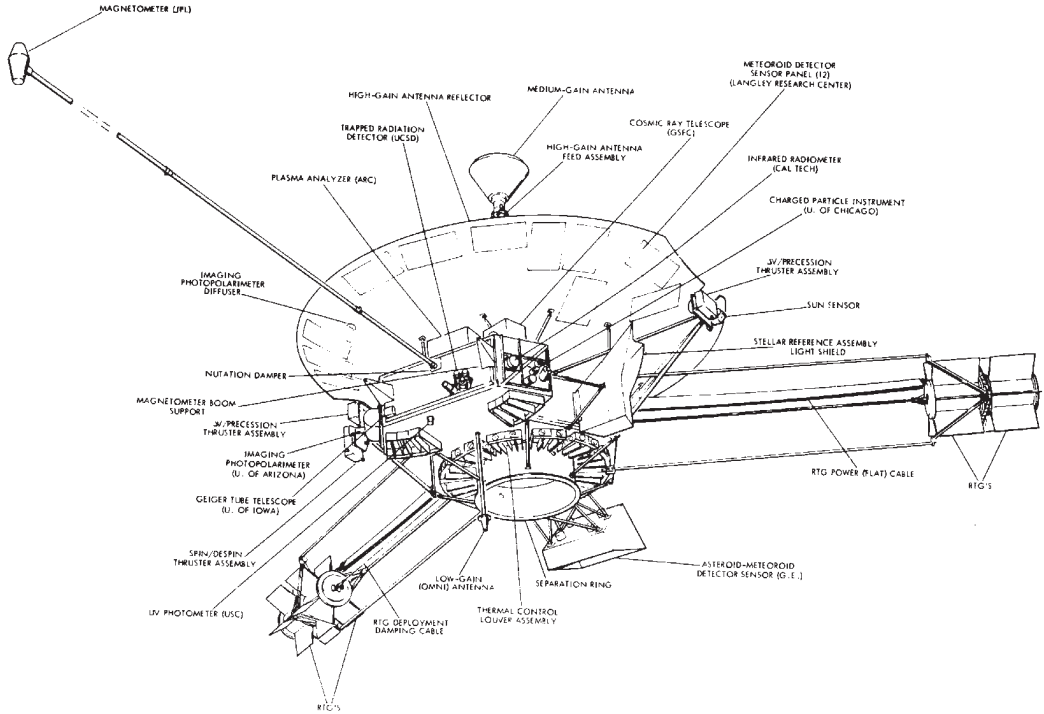

Figure 1: Schematic of the Pioneer spacecraft. From [1].

### III. THE PIONEER ANOMALY

Take an object like a spacecraft that is travelling through the solar system, influenced by the Newtonian gravitational potential of $n$ solar system bodies. If the object's initial position and velocity are known, its position and velocity at a later time can be calculated by solving the differential equations

$$\frac{d^2\mathbf{r}}{dt^2} = \sum_{i=1}^{n} \frac{\mu_i (\mathbf{r}_i - \mathbf{r})}{|\mathbf{r}_i - \mathbf{r}|^3}, \quad (1)$$

where the vector $\mathbf{r}$ is the position of the spacecraft, $\mathbf{r}_i$ is the position of the $i$th solar system body, and $\mu_i$ is its mass multiplied by Newton's gravitational constant. The equations can get somewhat more complicated if one incorporates nongravitational forces (e.g., solar pressure) or corrections due to general relativity, but the basic principle remains the same: with this set of equations and six numbers (the "initial state vector" describing the initial position and velocity), the motion of the object is fully characterized.

A signal travelling between a distant spacecraft and an Earth-based receiver would be altered by a Doppler frequency shift that is a function of the relative velocities of the transmitter and the receiver. The relativistic Doppler effect of an electromagnetic signal travelling from point $P_1$ (moving with velocity $v_1$) to $P_2$ (moving with $v_2$), transmitted with frequency $f_1$ and received with frequency $f_2$ can be calculated using the well-known formula

$$f_2 = f_1 \frac{(c - v'_2)\sqrt{c^2 - v_1^2}}{(c - v'_1)\sqrt{c^2 - v_2^2}}, \quad (2)$$

where $v'_1$ and $v'_2$ are the line-of-sight components (along $P_1P_2$) of the velocities $v_1$ and $v_2$ ($c$ is the speed of light).

To understand what happened to a signal, one needs to know when it was transmitted and received (so that, for instance, the motion of the transmitter and receiver at the appropriate times can be properly taken into account). An electromagnetic signal that travels through a gravitational potential field is delayed by that field; this phenomenon is known as the Shapiro time delay. In the gravitational field around a central body, the Shapiro time delay $\Delta t$ between two points $P_1$ and $P_2$, separated from the central body by distances $r_1$ and $r_2$, and from each other by the distance $r_{12}$, can be computed as [4]:

$$\Delta t = \frac{l}{c} \ln\left[\frac{r_1 + r_2 + r_{12} + l}{r_1 + r_2 - r_{12} + l}\right], \quad (3)$$

where $l = \mu(1 + \gamma)/c^2$, $\mu$ being the mass of the central body times the gravitational constant, and $\gamma = 1$[†].

Together, equations (1) through (3) can be used not only to calculate where a distant spacecraft is and how fast it is mov-ing, but also the frequency at which the spacecraft's signal is received by a ground-based receiver, the position and velocity of which is known to great precision. All we need is a set of six numbers: the initial state vector.

---

[†] For general relativity. For alternative gravity theories that can be described by what is known as the Parameterized Post-Newtonian, or PPN, formalism, $\gamma$ may have a different value.

These equations can also be used in reverse: if the Doppler frequency of the spacecraft's signal is measured at least six times, in principle we have six independent equations from which the six unknowns of the initial state vector can be determined. In practice, the equations are independent, but only barely so. That is because Doppler measurements are sensitive only to the line-of-sight velocity of the spacecraft, and as measurements are made across a time span measured in hours or days, the angular position of the spacecraft in the sky will change very little. Furthermore, the data can be noisy due to imperfections in the transmitter and receiver hardware, or to small forces and other effects that were not accounted for.

Therefore, instead of using only six measurements, a large number (several thousand or more) of Doppler measurements is used. The initial state vector is then found using a statistical method, for instance by least squares fitting. In practice, this approach works very well indeed: Doppler shifts amounting to a few milliHertz (mHz) can be measured (for a 2.2 GHz S-band radio signal, that's a relative precision of $10^{-12}$!) and the trajectory of the spacecraft can be determined very accurately.

When this process was followed for Pioneers 10 and 11, it was found that, after all known effects were taken into account and the trajectory of the spacecraft was calculated, the best statistical fit produced a result with a continuously varying Doppler shift. It is as if the spacecraft were slowed down by a small, constant force of unknown origin.

The most extensive analysis to date, performed by researchers at the Jet Propulsion Laboratory in 2002 [5], used about 11 years of Pioneer 10 and less than 4 years of Pioneer 11 data. It demonstrated unambiguously that the anomalous acceleration is present for both spacecraft. The magnitude of the acceleration is approximately $a_P = (8.74 \pm 1.33) \times 10^{-13}$ km/s$^2$, or about one tenth of one billionth ($10^{-10}$) of the gravitational acceleration on the surface of the Earth.

### IV. A Prosaic Explanation?

That two spacecraft flying in different directions produced consistent results strongly suggests that the anomalous acceleration is a real, physical effect. However, these two spacecraft have an identical design, and therefore, the possibility that the anomalous acceleration is of an on-board origin cannot be excluded.

Light carries momentum: $p = E/c$ ($E$ is the energy of a ray of electromagnetic radiation and $p$ is its momentum along the ray's direction). When you shine a flashlight in some direction, there is a tiny force pushing you in the opposite direction. The anomalous acceleration of the Pioneer spacecraft is so minuscule, light from a 65 W light bulb with a parabolic reflector would produce a comparable recoil force. Or not necessarily light: other forms of electromagnetic radiation will do just as well, including infrared radiation. In other words, heat.

On board the Pioneer spacecraft, the RTGs produce as much as 2500 W of heat, most of it radiated into space. The rest is converted into electrical energy that, in turn, also produces heat as it is used by the equipment on board. A small amount of electrical power is converted into radio energy, but that, too, would be pushing on the spacecraft in some direction as it is radiated by the antenna. Put all these thermal and radio emissions together, and a small anisotropy amounting to about 65 W of directed electromagnetic radiation is certainly not beyond the realm of the plausible. Nevertheless, previous studies suggested that there is not enough anisotropy in the electromagnetic radiation emitted by Pioneer 10 and 11 to produce the necessary force. However, no detailed analysis was performed of the spacecrafts' thermal radiation and its temporal evolution.

### V. Confirmation and Study of the Anomaly

There are two additional spacecraft escaping the solar system on hyperbolic trajectories: Voyagers 1 and 2. Unfortunately, these spacecraft are not spin-stabilized. The frequent use (several times a day) of thrusters on-board to maintain spacecraft orientation also introduces a significant amount of noise (thrusters being mechanical devices, the exact duration and strength of a thruster pulse is not known to great precision). Data from other spacecraft, such as Galileo, and Ulysses [3] were also analyzed, but no firm conclusions were reached. There were additional studies [6] attempting to establish a connection between the anomalous acceleration of the Pioneer spacecraft and what has become known as the "flyby anomaly", detected when several spacecraft flew by the Earth or other planets for a gravity assist ("slingshot") manoeuvre. These studies, too, remain inconclusive to date.

If only we knew the exact direction of the apparent acceleration, we could exclude many theories. A force of gravitational origin would presumably be dominated by the Sun, and thus it would point in the Sun's direction. A physical effect on the radio signal would produce an apparent acceleration along the line-of-sight direction. Interaction with the interplanetary medium (e.g., a drag force due to interplanetary dust) would produce a force along the direction of motion. Lastly, a force of on-board origin can be split into a component along the spin axis and a perpendicular component, the latter averaging to zero over time as the spacecraft spins; what remains is a force that, on average, points along the spin axis. Unfortunately these four directions closely coincide, and the data we have does not allow us to distinguish between them.

Absent any information about the exact direction of the anomalous force, no theories can be excluded. In the litera-ture, one finds papers that propose modified theories of gravity that would alter the spacecraft' motion; modified theories of electromagnetism that would have an effect on the frequency of the received signal, causing an apparent acceleration; and more exotic theories that, for instance, explore the possibility that clocks behave differently on board the spacecraft and on the Earth. Also, many noted the numerical coincidence $a_P \approx Hc$ where $H$ is Hubble's constant, although a cosmological theory would likely produce an acceleration pushing the spacecraft away from, not towards, the Sun.

Before we can decide which, if any, of these theories can be true, we need to develop a better understanding of the anomalous acceleration. One possibility is to launch a dedicated mission, or an instrument package on board another spacecraft. Yet another option is to use navigational data from NASA's New Horizons spacecraft, launched recently and now en route to Pluto. Even if these proposals are approved and funded, however, it would be many years before data is produced.

## VI. NEW ANALYSIS

There is something we can do today: analyze more data. There are nearly 20 years' worth of Pioneer 10 Doppler data and more than 15 years' worth of Pioneer 11 Doppler data that have never been used to check for the presence of an anomalous acceleration. Unfortunately, analyzing more data is not as simple as it sounds. A concerted effort last year recovered many sets of Doppler data [7, 8], but they come in several file formats on a variety of media, and the history of these files is uncertain; some contain raw readings, others already contain corrections. For instance, every revolution of the spinning spacecraft adds an extra cycle to the circularly polarized radio signal. At ~5 revolutions per minute, this means a difference of 0.2 Hz. While this does not sound like a great deal, it is quite significant when one searches for a discrepancy measured in mHz! Yet it is not evident which files already contain corrections for the spacecrafts' rate of spin and which do not.

Notwithstanding these difficulties, the effort continues and it is believed that complete and verified sets of Doppler data will soon be available for both Pioneers 10 and 11, covering their entire mission durations from launch until the last data point.

Doppler data alone is not sufficient for precision orbit determination. One also needs to know what kinds of manoeuvres were performed and when; additionally, knowledge of the spacecrafts' exact spin rate is important. While the missions were under way, such information was extracted from the telemetry data stream. Much to our fortune, the telemetry data received from Pioneers 10 and 11 have been preserved. This has more to do with luck than design: tapes containing telemetry files were scheduled to have been destroyed years ago, but there was no funding to do so.

Today we are in possession of telemetry files that contain approximately 40 gigabytes of data; nearly every bit transmitted by these spacecraft and received by stations on the Earth[‡]. Making sense of these telemetry files was akin to a detective job. Indeed, much of the current research is like uncovering clues from an old crime scene, as it involves first locating, and then interpreting 30-40 year old design documents and operational notes, some of which are hand-written scraps of paper.

As a result of our work, we now have software tools that can be used to determine the physical state of both spacecraft at any time during their journeys. In particular, we know the time and type of each manoeuvre that was performed; changes in the spacecraft spin rate; performance of on-board electrical systems; and equipment and electronics platform temperatures. Telemetry records even tell us about the received signal strength and quality at the DSN station. Together, the recently recovered Doppler and telemetry files allow us to virtually "re-fly" both Pioneer missions, analyzing their trajectories and testing new hypotheses. We hope that, by the time our work is completed, we will be able to firmly decide if the anomalous acceleration can be due to forces generated on-board.

Recovering the history of a space mission that spanned more than 30 years is an unprecedented undertaking. The use of telemetry data for orbital analysis, to our knowledge, was never attempted previously. The task at hand requires a thorough understanding of a spacecraft launched almost 35 years ago, and ground data systems and software developed and deployed in the 1960s. These are just some of the unique aspects of our present work, conducted together with colleagues from several countries, as part of an international collaborative effort [9] to resolve the Pioneer Anomaly.

---

[‡] Some files are missing due to damaged or lost tapes. Most notably, we do not have telemetry for several days during Pioneer 10's Jupiter encounter, and longer stretches of data are missing from later parts of Pioneer 11's mission.